\documentclass{easychair}

\usepackage{doc}
\usepackage{multicol}
\usepackage{relsize}
\usepackage{float}
\usepackage{subcaption}
\usepackage{graphicx}

\hypersetup{hidelinks}

\title{Learned Clause Minimization in Parallel SAT Solvers\thanks{This work 
received funding from the BMBF under grant 01IH15006C 
(HPSV).}}

\author{
Marc Hartung\inst{1}
\and
Florian Schintke\inst{1}
}

\institute{
  Zuse Institute Berlin,\\
  Berlin, Germany\\
  \email{\{hartung,schintke\}@zib.de}
 }

\authorrunning{M. Hartung and F. Schintke}

\titlerunning{Learned Clause Minimization in Parallel SAT Solvers}

\begin{document}

\maketitle

\begin{abstract}
Learned clauses minimization (LCM) led to 
performance improvements of modern SAT solvers especially in solving hard SAT 
instances. Despite the success of LCM approaches in sequential solvers, they 
are not widely incorporated in parallel SAT solvers. In this paper we explore 
the potential of LCM for parallel SAT solvers by defining several LCM 
approaches based on clause vivification, by comparing their runtime in 
different 
SAT solvers, and by discussing reasons for performance gains and losses. 
Results 
indicate that LCM only boosts performance of parallel SAT solvers on a fraction 
of 
SAT instances. Applying LCM more commonly decreases performance. Only certain 
LCM approaches are able to improve the overall performance of parallel SAT 
solvers.
\end{abstract}

\section{Introduction}
In recent years more modern Conflict-Driven Clause Learning (CDCL) SAT solvers 
incorporate learned clause minimization (LCM) to increase their overall 
performance. Despite several clause minimization (CM) approaches used in 
SAT solvers, LCM mostly refers to vivification~\cite{piette_vivifying} of 
learned clauses, where redundant literals are removed from a clause by 
unit propagation and conflict analysis. The performance 
increase results from a more efficient propagation and a higher chance of 
producing unit clauses due to shortened clauses~\cite{luo_an_effective}.
The success of LCM in modern SAT solvers is supported by the results 
of the main track of the SAT Competition'18, where the thirteen highest 
ranked solvers all use LCM. However, this success seems not to translate 
directly to 
parallel SAT solving. The highest ranked LCM solver of the parallel 
track is ranked third and only a fraction of the track's participants 
use 
LCM at all.
Parallel SAT solvers commonly are portfolio solvers, whereby multiple 
sequential CDCL SAT solver instances run different heuristics in parallel. 
Each solver instance exports presumably beneficial learned 
clauses to other instances. Another parallel approach is 
divide-and-conquer based SAT 
solving~\cite{nejati_propagation,Heule2018,audemard_effective,
hyvarinen_partitioning}, which will not be covered in this paper. 

In general, the challenge of incorporating LCM is choosing
the trade-off between time spent on CDCL solving and on 
minimizing clauses. Even in sequential solvers this remains challenging due to 
the diversity in structure of SAT problems. In parallel solvers this challenge 
seems to 
be even greater. Probably, reasoning on a variety 
of imported beneficial clauses leads to more relevant conflict clauses, which 
then can be exported again. This creates a leverage effect, whereby spending a 
significant time on LCM may slow down the overall progress, because most 
shortened clauses are only used by a fraction of solvers.

In this paper we analyze homogeneous LCM approaches, i.e., every solver 
instance of a parallel SAT solver uses the same clause minimization approach 
instead of dedicating a subset of instances to clause minimization. This has 
two reasons: First, heterogeneous solver structures need to be load 
balanced~\cite{wieringa_concurrent}. It has to be ensured that all chosen 
clauses are minimized without stalling any thread, which becomes more difficult 
with increasing parallelism. Second, successfully minimizing a clause is more 
likely, when the solver that learned this clause minimizes it. This solver has 
a higher probability of holding additional conflict clauses that cut similar 
parts of the search tree.

Our main contributions are:
\begin{itemize}\setlength{\itemsep}{0pt}\setlength{\parskip}{0pt}
\item A general overview of applicable homogeneous LCM approaches in parallel 
SAT solvers (Section~\ref{sec:cm_parallel}),
\item an extended versions of Glucose-Syrup 4.0 implementing the proposed LCM
approaches (Section~\ref{sec:viv_syrup}), and
\item an extensive performance comparison of Glucose implementations and 
the vivification SAT solvers TopoSAT2 and Sticky 
(Section~\ref{sec:other_solvers}).
\end{itemize}

\section{Background}
In the following, we assume knowledge of fundamental techniques of SAT
solving, such as Boolean constraint propagation (BCP), backtracking,
clause learning and resolution, as well as basic parallel approaches
for SAT solving, especially portfolio solvers with clause sharing. We
cover clause minimization approaches only briefly and refer to given
references for more details.

Clause minimization is applied either during preprocessing or in-proccessing. 
Common preprocessing techniques are clause and variable elimination based on 
SatElite~\cite{een_effective}. Thereby, variables are eliminated through 
resolution~\cite{Subbarayan_niver} and redundant clauses are removed using 
(self)-subsumption until the given formula cannot be further minimized. This 
approach is not limited to preprocessing, but the impact is said 
to be not significant in most cases~\cite{Biere_preprocessing}.

Multiple in-proccessing techniques were already proposed. For example, most 
Minisat~\cite{een_extensible} based solvers apply recursive clause 
minimization~\cite{Sorensson_minimizing} and 
binary resolution to remove literals from currently generated conflict clauses.
More recently, serial SAT solvers incorporate clause vivification to minimize 
initial and learned clauses. Vivification is also known as distillation and was 
first proposed for incremental SAT solving~\cite{JIN200551,Han_alembic} and as 
preprocessing approach~\cite{piette_vivifying}. Later, it was used for 
in-processing~\cite{luo_an_effective}, which is the fundamental approach for 
our analysis of in-processing clause minimization in parallel SAT solvers.
Thereby, the negation of each literal of a clause is propagated. During 
propagation three cases can occur: 
(1) When a conflict is detected, the resulting conflict clause replaces the 
original clause. (2) A literal is propagated to true, whereby not yet 
propagated literals can be removed from the clause. (3) Literals that are 
non-decisionally propagated to false, can be removed from the clause. 
In general, vivifications are only independent from the search space, when 
applied at decision level zero and therefore are only applicable after restarts 
or complete backtracks. Due to the large overhead of applying unit 
propagation and the probability that no literal can be removed, vivification 
is only applied to a subset of clauses. A possible subset are clauses that 
are most likely to be kept in the near 
future, i.e., clauses with a low LBD value~\cite{audemard_glucose} 
or activity~\cite{audemard_satcomp18}.

Approaches of clause minimization in the context of parallel SAT solving are 
barely studied. SArtAgnan~\cite{Kaufmann_sartagnan} dedicates one of eight 
threads to minimize all initial clauses and a subset of learned clause using 
the SatElite approach. Another approach~\cite{wieringa_concurrent} uses one 
reducer thread besides a CDCL solver. The reducer performs vivification on 
learned clauses and communicates successful minimizations to the CDCL solver. 
Two approaches of the SAT Competition '18 use a homogeneous approach, where 
every thread tries to minimize learned clauses. Thereby, ABCDSat applies a 
variation of vivification to learned clauses with 
small LBD value and removes redundant clauses by a subsumption 
check~\cite{chen_satcomp18}. In contrast, TopoSAT2 vivifies clauses before 
export~\cite{ehlers_satcomp18}. However, no studies on 
such homogeneous approaches are available and their impact is not evaluated 
yet. Due to the lack of insights and publications, we are not able to discuss 
Cryptominisat~\cite{soos_extending}, which also uses vivification during active 
solving.

\section{Clause Minimization in Parallel}\label{sec:cm_parallel}
\begin{figure}[b]
  \begin{tabular}{p{0.475\textwidth}p{0.475\textwidth}}
\vspace{0pt}\includegraphics[width=\linewidth]{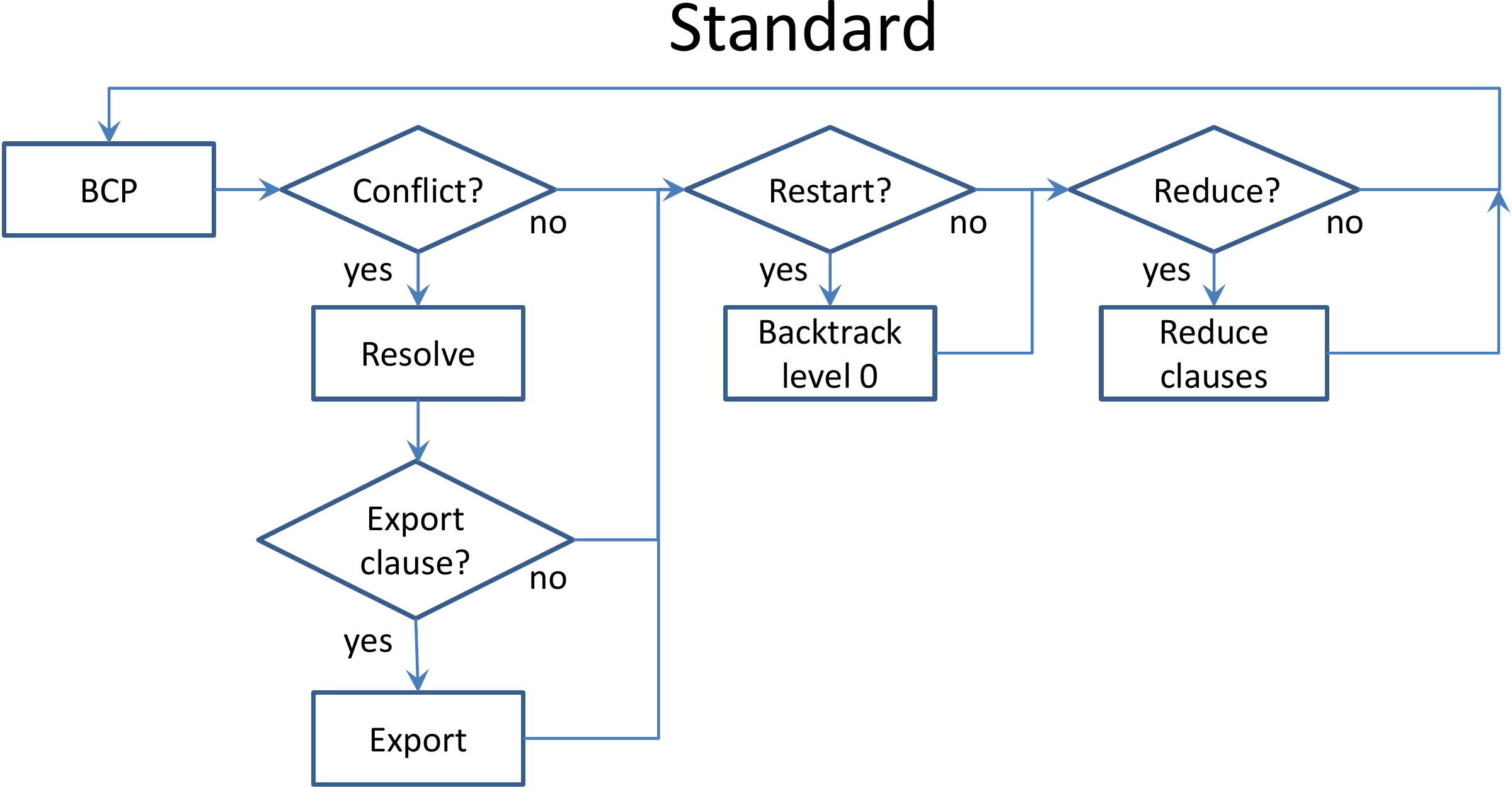}&
\vspace{0pt}\includegraphics[width=\linewidth]{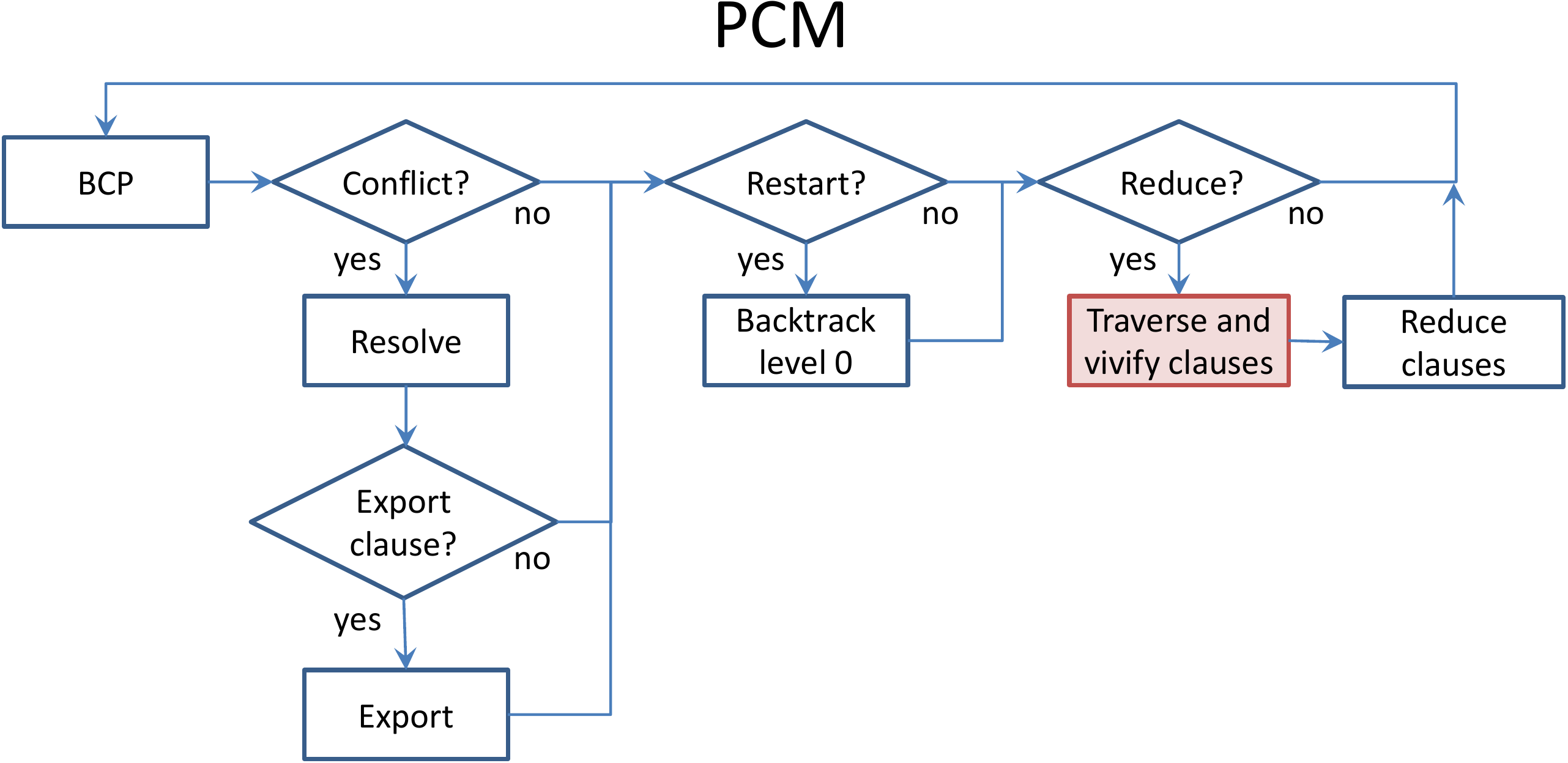}\\
\vspace{0pt}\includegraphics[width=\linewidth]{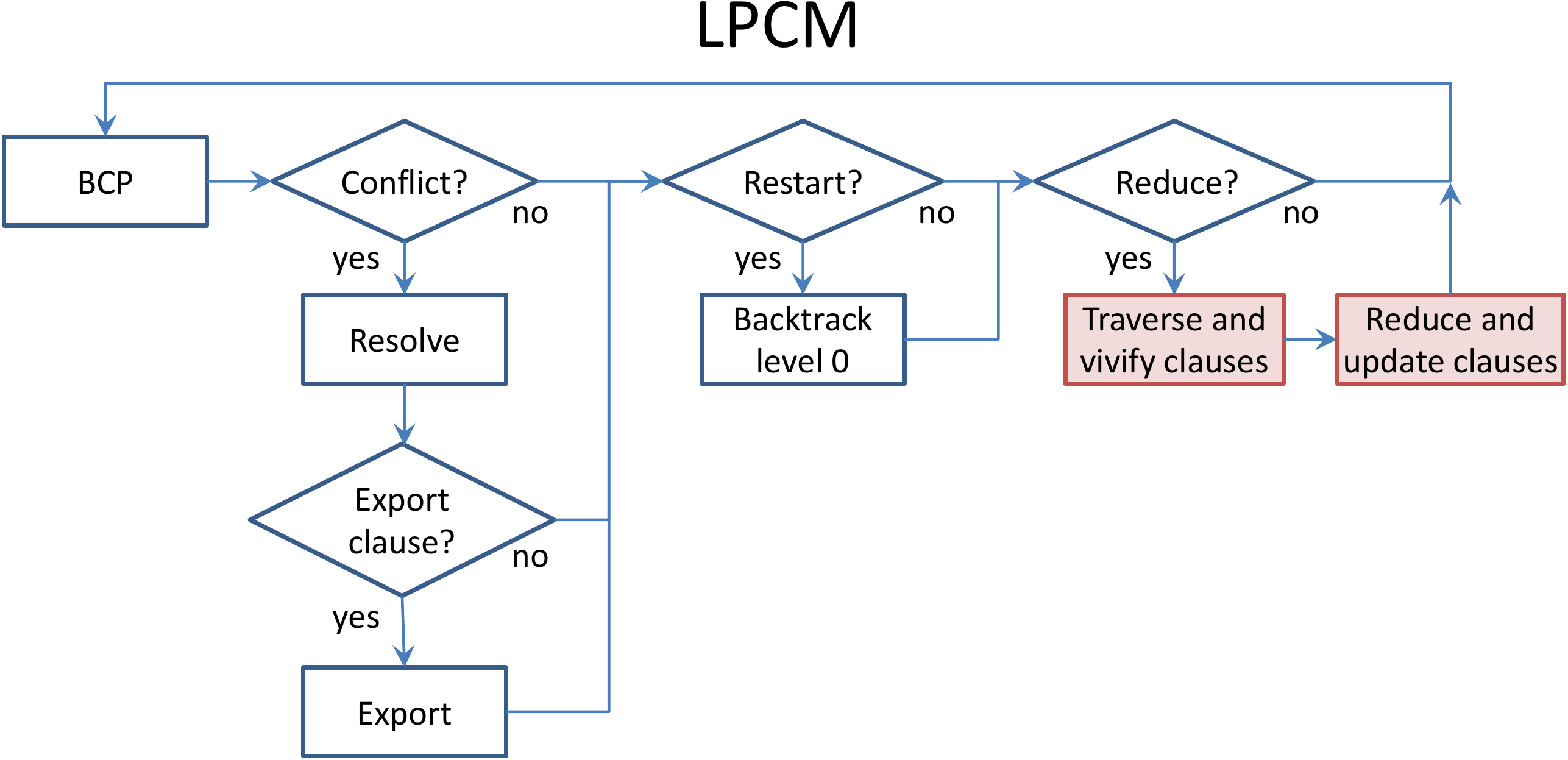}&
\vspace{0pt}\includegraphics[width=\linewidth]{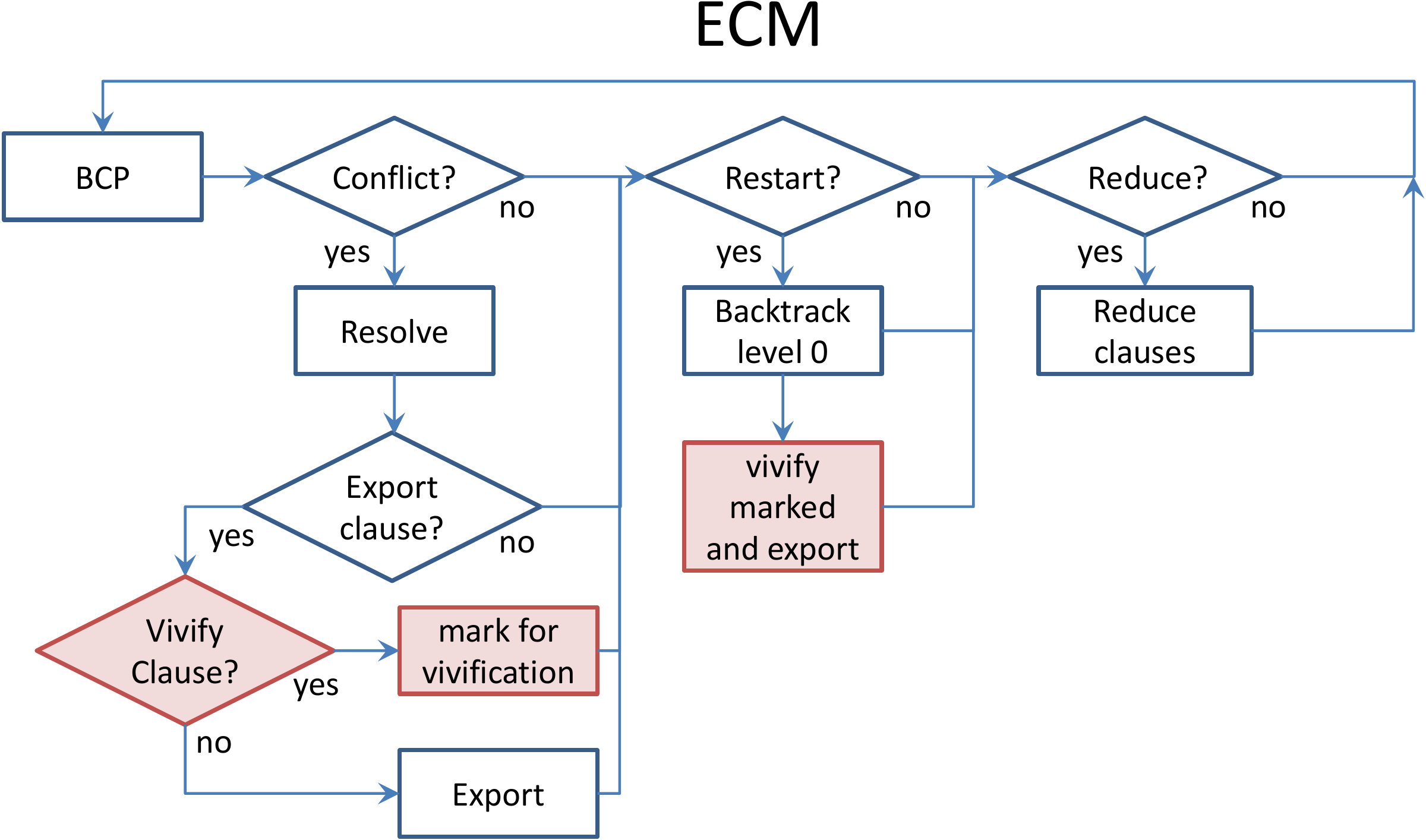}
  \end{tabular}
\caption{The flowcharts show the workflow of each proposed LCM approach. The 
standard workflow depicts the basic CDCL components of modern parallel SAT 
solvers.}  
  \label{fig.flow-charts}
\end{figure}

The basic LCM approach described in the previous section is trivially 
portable to most modern parallel SAT solvers. However, sharing improvements of 
clauses is challenging due to redundant clause storage. We describe 
three homogeneous clause vivification approaches, which target this challenge.
In Figure~\ref{fig.flow-charts} the differences between the approaches are 
outlined using flowcharts. For comparison, the standard CDCL workflow is given 
and necessary changes for incorporating each proposed approach are marked 
in red.

\paragraph{Private Clause Minimization.}
Private clause minimization (PCM) resembles the common learned clause 
minimization~\cite{luo_an_effective}, whereby only learned clauses are 
minimized, when they are likely to be kept. In context of parallelism, PCM 
solvers only minimize clauses that were learned by them, i.e., do not vivify 
imported clauses. In an early stage of
our research we discarded vivification of imported clauses due to two insights 
we gained: (1) Vivification of imported clauses fail more often, most likely
caused by different search spaces of exporting and importing solvers, 
and (2) the number of kept clauses increases heavily, wherefore the heuristic 
cut on which clauses are chosen for vivification has to be more restrictive to 
prevent an imbalance between CDCL solving and LCM. This leads to a
worse performance than disabling LCM in most cases. The vivification
is always triggered 
before the clause database is reduced. For example, in Glucose-Syrup it is 
therefore necessary to delay the clause reduction until the solver reaches 
level 0 or the current implication graph has to be stashed to be able to apply 
vivification. Since vivification is triggered independently from the export 
routine, clause improvements are never exported, when no lazy export policy is 
used~\cite{audemard_lazy}.

\paragraph{Linked PCM (LPCM).}
To overcome the issue of unshared shortened clauses, each copy of an exported 
clause is linked by a unique reference in LPCM. Thus, the learning solver can 
vivify a 
clause and on success can share the strengthened clause by using the link. For 
the 
linked PCM (LPCM) implementation of Syrup we added an atomic pointer to each 
exported clause that points to a reference counted array containing the new 
clause. Then, importing solvers can check during clause reduction,
whether an improvement of a clause is present. This scheme can be easily 
applied to most SAT solvers. In contrast, the SAT solver 
Sticky uses physical clause sharing~\cite{wieringa_concurrent}, wherefore such 
links are already in place, since clauses are not copied during sharing. 

\paragraph{Export Clause Minimization.}
Instead of sharing a clause's improvement after the clause was exported, 
exports can be delayed until the solver is able to vivify the clause, so that 
only the improved clause is shared. This vivification approach, we
call it Export 
Clause Minimization (ECM), was first proposed and 
implemented by Ehlers et al.~\cite{ehlers_satcomp18} in the solver TopoSAT2. 
Thereby, it is insufficient to vivify every exported clause, due to a potential 
high workload on the minimization. Vivification should be restricted by an 
LBD~\cite{ehlers_tuning} or size limit. Two theoretical disadvantages occur due 
to this workflow. First, the export of clauses with good LBD values can be 
significantly delayed, which can lead to a performance decrease. Second, all 
clauses with a sufficiently low LBD are vivified, which can lead to a high 
overhead due to vivification. Our experiments show that ECM should be 
restricted to clauses with an LBD lower than 4.

\section{Experimental Results}
Optimizing heuristics in parallel SAT solvers is heavily compute-intense. Due 
to restrictions in compute time, we focus on Glucose based solvers to save 
initial optimizations on which parameters work well with LCM and ECM.

We implemented PCM, LPCM and ECM in Glucose-Syrup (Syrup) and 
further compare it to TopoSAT2, which implements ECM, and to our experimental 
SAT solver Sticky, which implements LPCM and ECM. The benchmarks are taken from 
the main tracks of the SAT Competition 2018 and 2017 as well as the application 
track of the SAT Competition 2016. All tests ran on single socket Intel Xeon 
Phi 7250 with 68 cores at 1.4\,GHz clock rate and 96\,GB main memory, whereby 
the MCDRAM was disabled. Due to memory restrictions each solver only uses 34 
solver instances in parallel. For each benchmark a time limit of 15,000 seconds 
was set. The time limit is set three times higher than in the SAT Competition, 
because of the lower clock rate and the slower memory subsystem of the used
hardware. All results and the Syrup implementations are available 
online\footnote{\url{https://github.com/marchartung/glucose4_0_vivi}}.

\subsection{Vivification in Syrup}\label{sec:viv_syrup}

The Syrup extensions are based on the first 
parallel version 4.0 of Glucose. The VSIDS branching~\cite{moskewicz_chaff} and 
the dynamic restart~\cite{audemard_refining} heuristics as well as the clause 
database management are untouched in all extensions. In the PCM implementation 
vivification is triggered before the clause database is reduced, wherefore 
reduction may be delayed until the solver restarts. For identifying clauses 
worth minimizing, all clauses are sorted according to their LBD and activity 
value and only the half with the lowest LBD values and at most an LBD of 5 or 
lower are vivified. The solvers only try to vivify a clause once. As mentioned 
before, minimized clauses are not shared again. LPCM is enabled by reserving 
extra space for an atomic pointer in each shared clause. When a clause is 
shared before the solver tried to vivify it, the pointer is set to an 
array, where the vivified clause is copied to later. Every solver can 
then check during reduction whether a clause was minimized. The rest of the 
minimization process of LPCM is equal to PCM. The ECM implementation delays 
exports of clauses with an LBD lower than 4 (syrup-ECM3) or 5 (syrup-ECM4) 
until a restart occurs and then vivifies and exports them. These clauses will 
not be removed before the vivification took place and thus are protected 
during the database reduction process.

The following factors have to be considered to reason about performance 
differences between implementations. The two major performance decreasing 
factors are 
the slowdown caused by the propagations spend for the clause vivification and 
the overhead introduced by sharing the clause improvements. Both decreasing 
factors are only justified, when minimized clauses sufficiently cut the search 
space or participate in the final proof, i.e., significantly shorten the time 
to 
solution.

\begin{figure}
    \centering
    \begin{subfigure}[h]{0.495\textwidth}
        \caption{SAT}
        \includegraphics[width=\textwidth,clip,trim={78 48 70
140}]{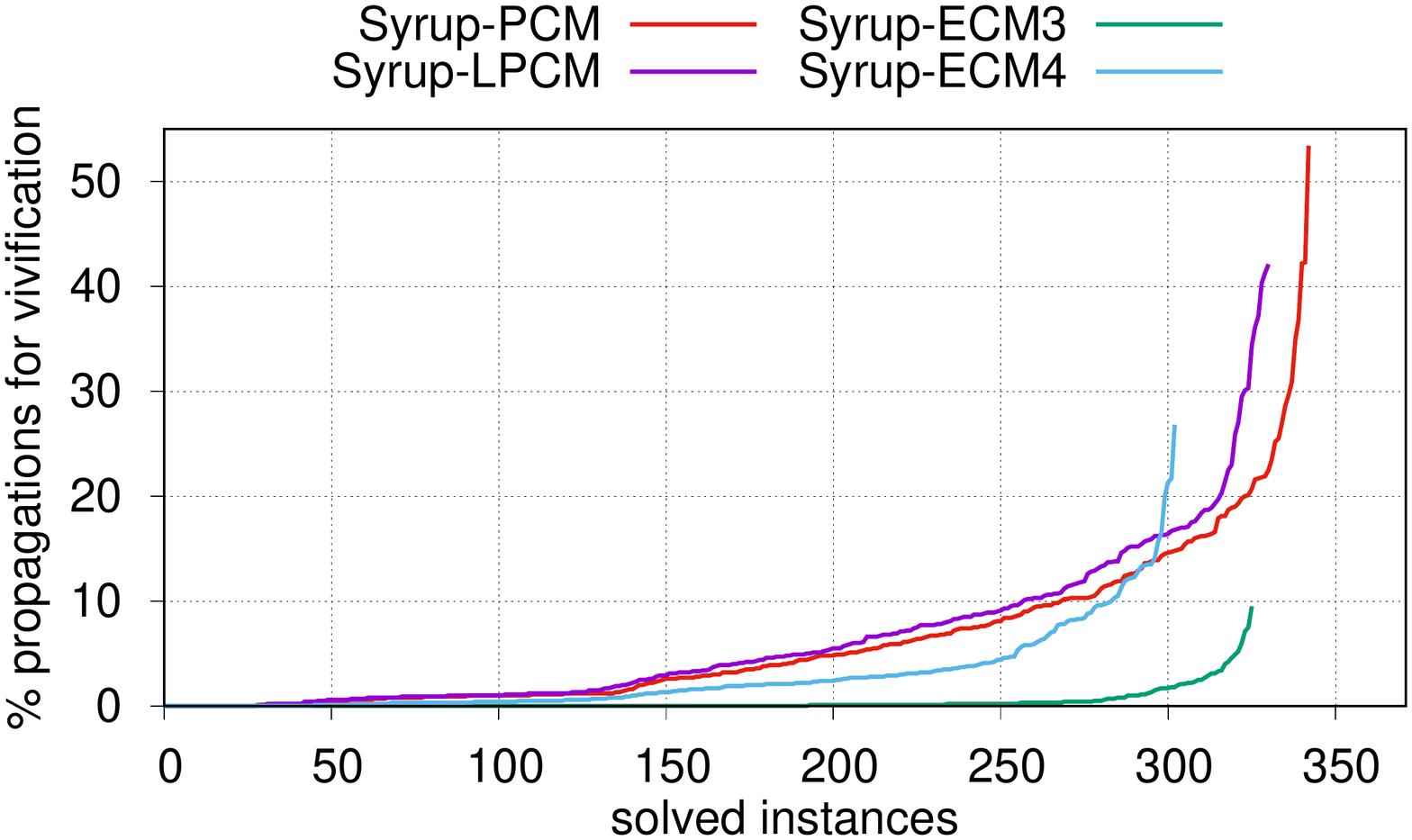}
        \label{fig:vprops_sat}
    \end{subfigure}
    \begin{subfigure}[h]{0.495\textwidth}
        \caption{UNSAT}
        \includegraphics[width=\textwidth,clip,trim={50 48 110 
140}]{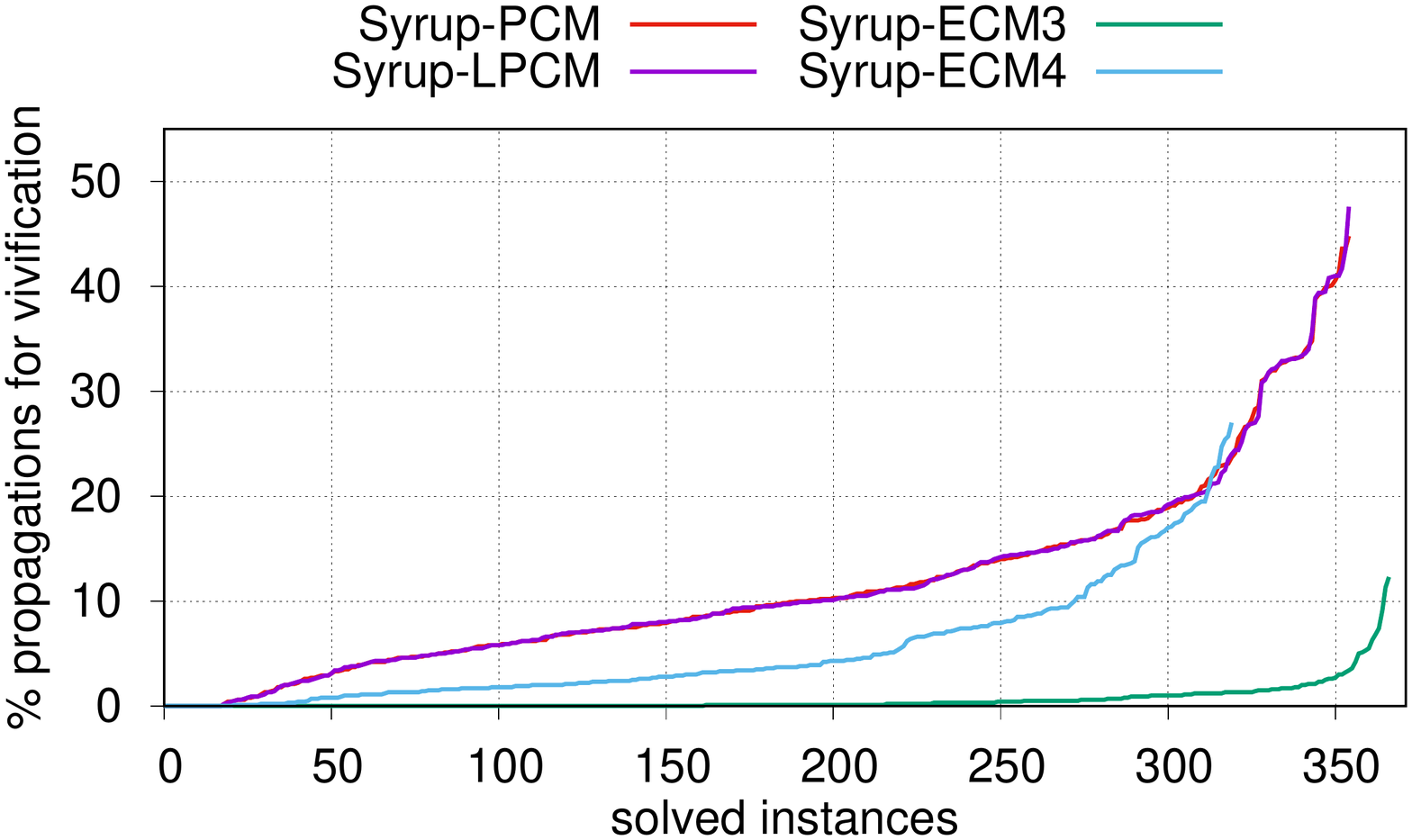}
        \label{fig:vprops_unsat}
    \end{subfigure}
    \caption{Percentage of propagations used for clause vivification by each 
approach.} 
\label{fig:cactus_vprops}
\end{figure}

To estimate the vivification overhead, Figure~\ref{fig:cactus_vprops} shows 
cactus plots of the percentage of propagations spent for 
vivification of each Syrup implementation on the solved SAT (left) and UNSAT 
(right) instances. ECM3 spends less than 1\,\% on 
propagations for vivification in over 250 SAT instances and in nearly 300 UNSAT 
benchmarks, which is far less compared to ECM4, where only about 50 SAT and 
UNSAT instances have such a low vivification overhead. CDCL solvers have a 
higher chance to generate clauses with a higher LBD and clauses with a higher 
LBD tend to have more literals. Thus, the increase of vivification overhead 
from ECM3 to ECM4 is plausible. PCM and LPCM, both spent roughly the
same amount of more propagations for vivification than the ECM approaches.

\begin{figure}
    \centering
    \begin{subfigure}[h]{0.495\textwidth}
        \caption{SAT}
        \includegraphics[width=\textwidth,clip,trim={20 0 0 0}
        ]{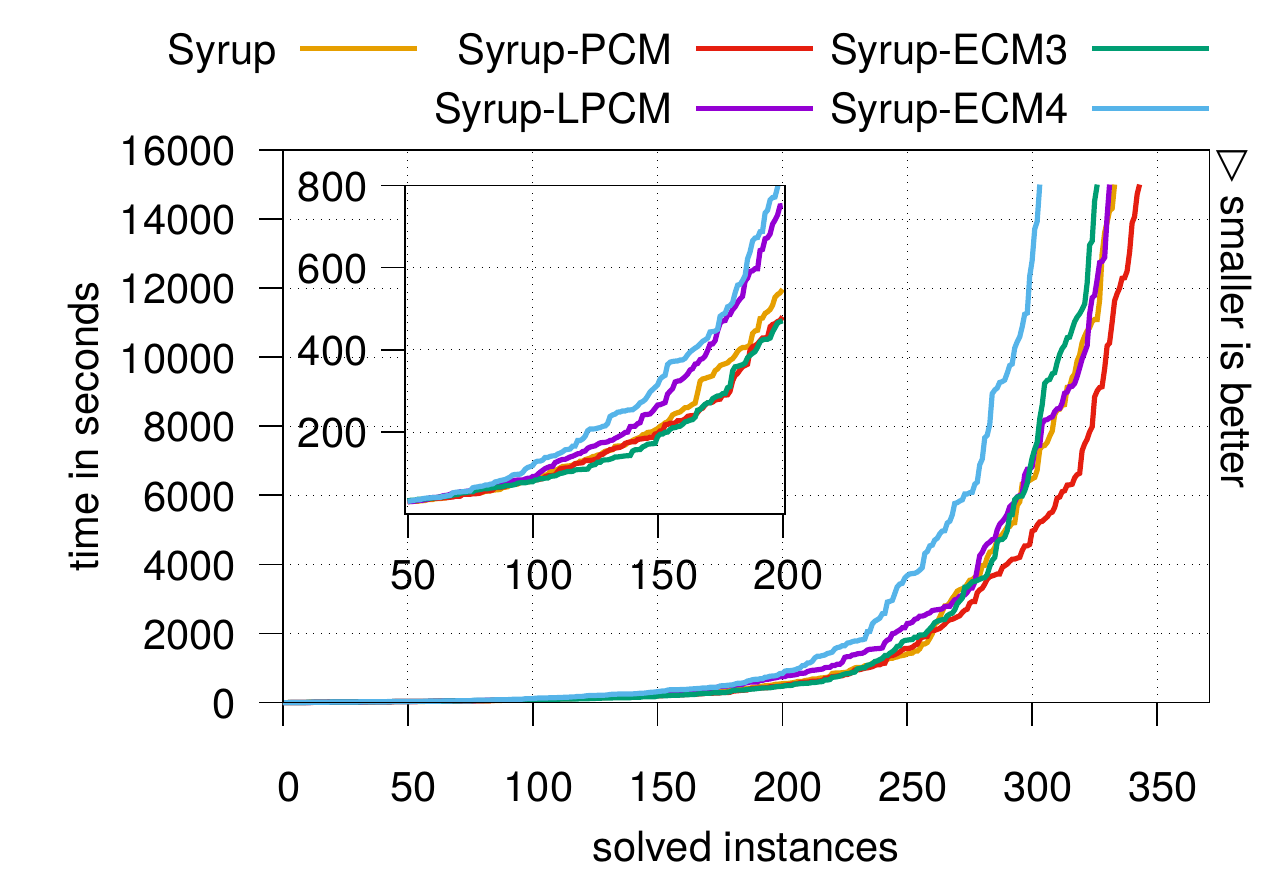}
        \label{fig:runtime_sat}
    \end{subfigure}
    \begin{subfigure}[h]{0.495\textwidth}
        \caption{UNSAT}
        \includegraphics[width=\textwidth,clip,trim={0 0 0 0}
        ]{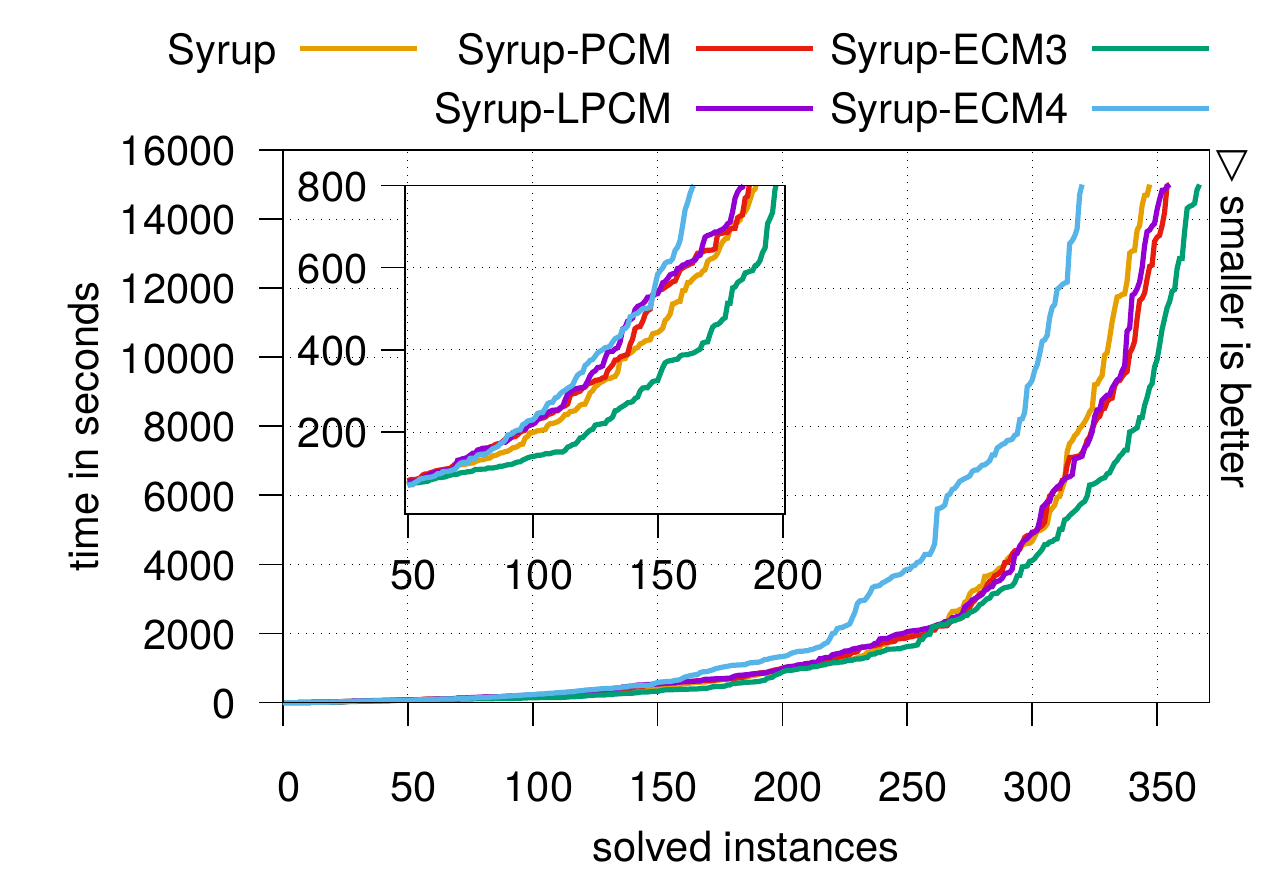}
        \label{fig:runtime_unsat}
    \end{subfigure}
    \caption{Runtime cactus plot of solved instances by every Syrup 
implementation.} 
\label{fig:cactus_runtime}
\end{figure}

\begin{figure}
    \centering
    \begin{subfigure}[b]{0.49\textwidth}
        \caption{SAT}
        \includegraphics[width=\textwidth,clip,trim={0 5 200 
50},clip]{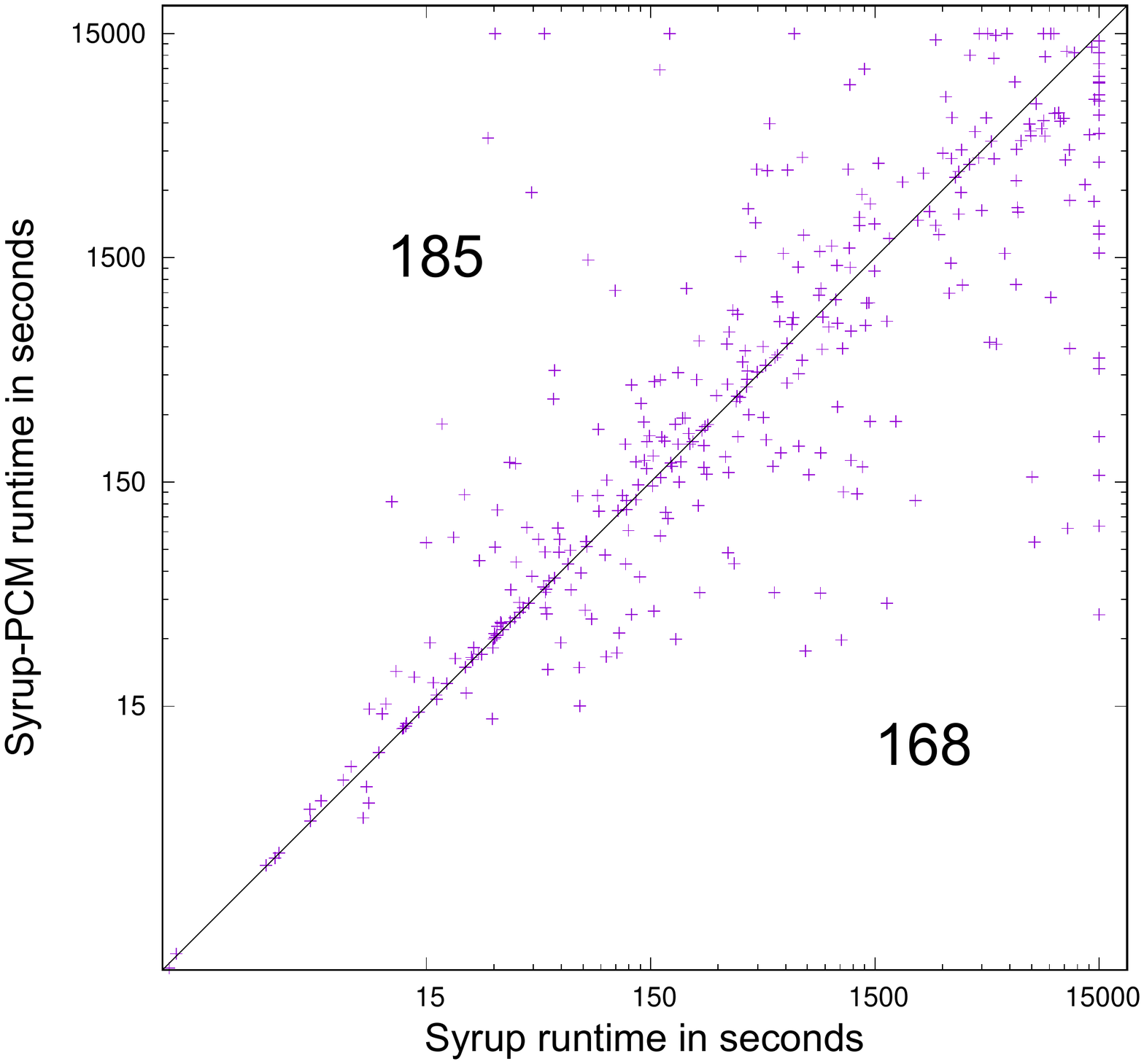}
        \label{fig:pcm_sat}
    \end{subfigure}
    \begin{subfigure}[b]{0.49\textwidth}
        \caption{UNSAT}
        \includegraphics[width=\textwidth,clip,trim={0 5 200 
50},clip]{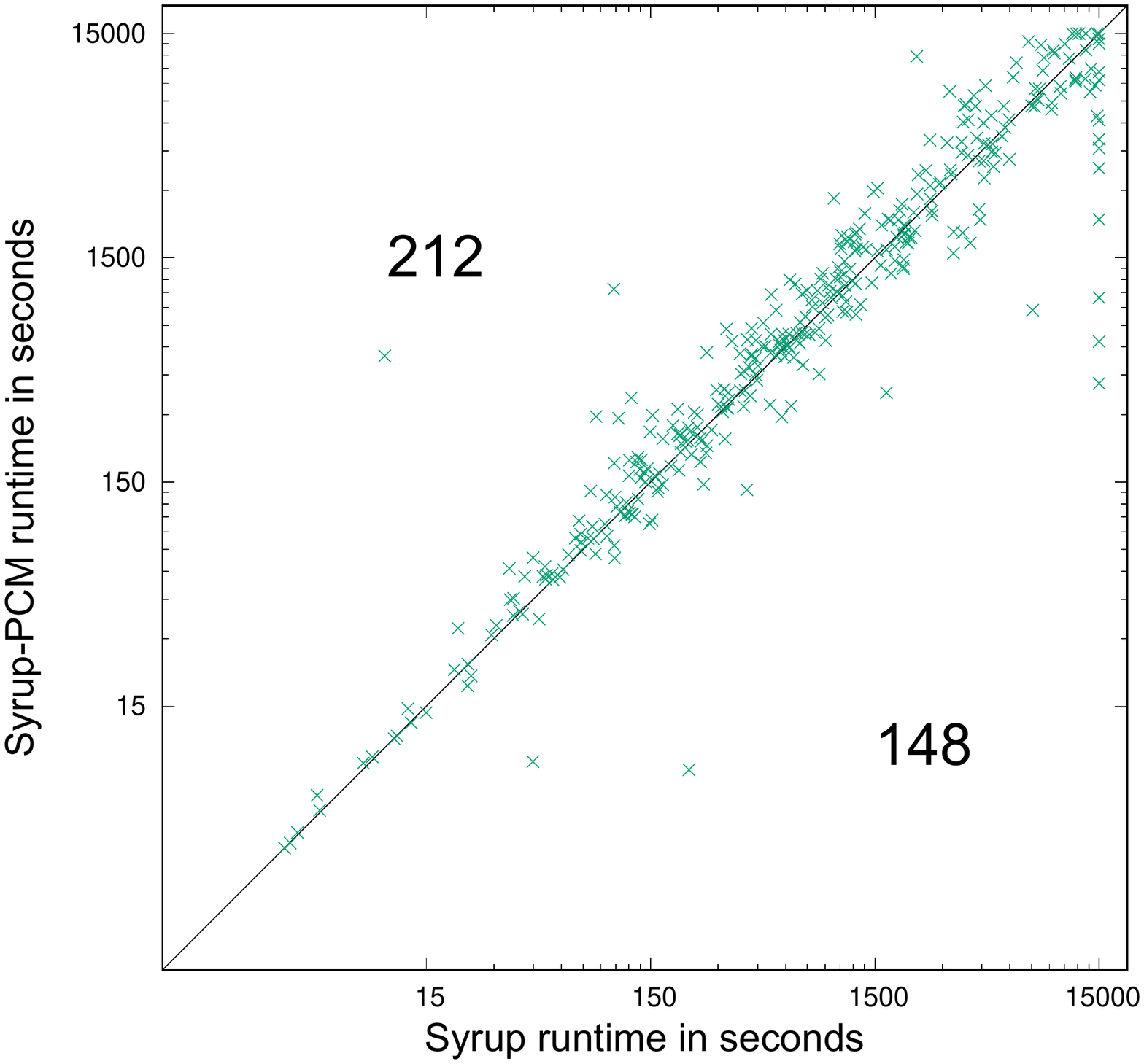}
        \label{fig:pcm_unsat}
    \end{subfigure}
    \begin{subfigure}[b]{0.49\textwidth}
        \caption{SAT}
        \includegraphics[width=\textwidth,clip,trim={0 5 200 
50},clip]{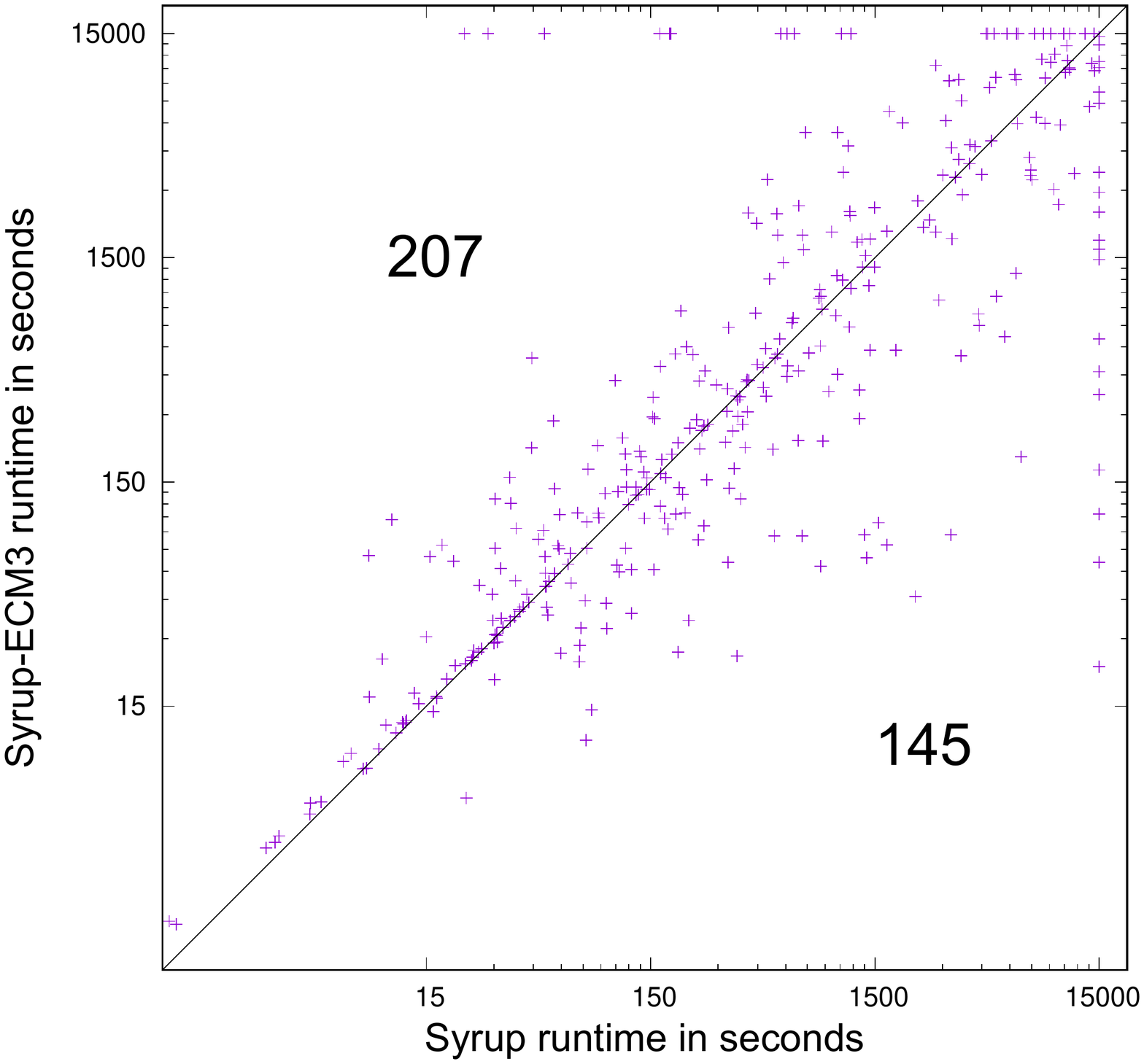}
        \label{fig:ecm_sat}
    \end{subfigure}
    \begin{subfigure}[b]{0.49\textwidth}
        \caption{UNSAT}
        \includegraphics[width=\textwidth,clip,trim={0 5 200 
50},clip]{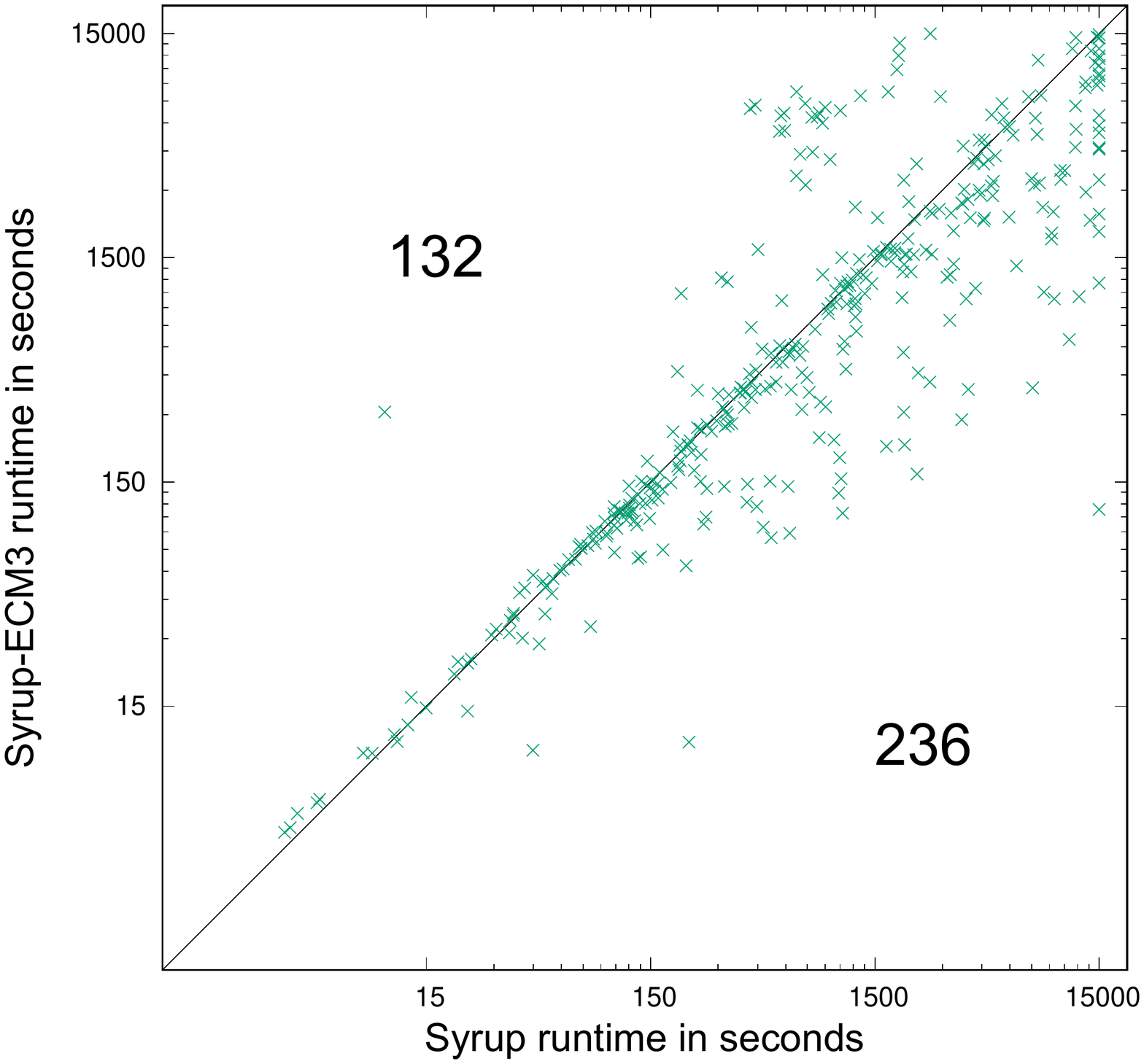}
        \label{fig:ecm_unsat}
    \end{subfigure}
    \caption{Runtime comparison plots of Syrup to Syrup-PCM (a, b) and Syrup 
to Syrup-ECM3 (c, d). A point below the diagonal means Syrup was faster. The 
numbers in the upper and lower triangles determine the number of points in the 
triangle.} 
\label{fig:runtime}
\end{figure}

The time spent on vivification is not an indicator for its impact. Even minor 
clause improvements can significantly improve the solver's performance. An 
example for this is the performance decrease from ECM3 to ECM4 in 
Figure~\ref{fig:runtime}, which shows the runtime distribution of the Syrup 
implementations, whereby an additional zoom plot is given for the range between 
50 and 250 solved instances. In the satisfiable and unsatisfiable case, ECM3 
is always faster than ECM4 despite seldom but very effective clause 
minimizations. Also, the percentage of successful shortened clauses over all 
solved instances is with 6.2\,\% for ECM3 much lower than for ECM4 with 
32.6\,\%.
The impact of vivification depends much more on how often vivified clauses 
are reused by the solvers. Audemard et al.~\cite{audemard_lazy} state that most 
of the learned 
clauses participate in one conflict, but only a small percentage appears in 
multiple conflicts (91\,\% appear one time, 34\,\% two times, 22\,\% three 
times, 
17\,\% four times). Therefore, vivifying clauses with a high 
activity is more likely to have a higher impact due to higher probability of 
reusage. This assumption is substantiated by the results of PCM and LPCM, which 
prefer vivifying active clauses, where the vivification overhead is larger than 
in any other approach (both have a success rate of $\approx$39\,\%), but 
compared 
to ECM4 there is an increase in solved instances. Additionally, it is stated by 
Audemard et al.~\cite{audemard_lazy} that on average only 10\,\% of
the imported clauses lead to 
conflicts in the importing solver. Thus, heavily vivifying export clauses has 
minor impact for importing solvers, since most of the clauses will not 
be used.

Regarding the changes in performance through vivification to the default Syrup 
implementation, ECM4 decreases the number of solved instances and the time to 
solution for SAT as well as UNSAT instances. As discussed above, the 
vivification overhead seems to be larger than the vivification impact.
Contrary, PCM increases the number of solved instances on SAT and UNSAT, which 
is visible in Table~\ref{tab:complete} more clearly, where the number of solved 
instances of all evaluated vivification solvers are displayed. Thereby, the 
increase in solved instances (3 additional instances for SAT'16A and 18 for 
SAT'18) is comparable to the increase of the LCM approach in 
Glucose~\cite{luo_an_effective} (11 additional instances for SAT'14A and 6 for 
SAT'16A). The additional overhead of LPCM through managing additional resources 
for linking clauses, seems to cancel out the positive effects of 
vivification for SAT instances. LPCM solves two fewer SAT instances than the 
default implementation over all SAT Competitions and does not solve more UNSAT 
instances than PCM. The same scenario occurs much more clearly, when using 
ECM3, where 7 fewer SAT instances and 20 more UNSAT instances are
solved compared to the default implementation.

Considering the runtime behavior, the default implementation solves
185 SAT instances faster than its PCM implementation and PCM solves
only 168 instances faster (see Figure~\ref{fig:pcm_sat}). For the
UNSAT instances the difference is with 212 to 148 even greater (see
Figure~\ref{fig:pcm_unsat}). But on average this translates only to a
runtime increase of less than 2\,\% over all solved
instances. Regarding ECM3, the overall runtime is shortened by
$\approx$4\,\%. In Figure~\ref{fig:runtime}(c) and (d) it is visible
that this runtime improvement is mostly based on an improved
performance in unsatisfiable instances.

\subsection{Comparison to other SAT solvers}\label{sec:other_solvers}
In the following, we compare our Syrup implementations with TopoSAT2 and 
Sticky. 
Both solvers incorporate vivification as an additional clause minimization 
approach, whereby TopoSAT2 uses ECM and Sticky is capable of using ECM or LPCM.

\paragraph{TopoSAT2} has an optional ECM feature similar to our ECM 
implementation in Glucose. The main difference is that clauses are not 
protected during reduction. Instead, all clauses with an LBD lower than 4 are 
copied to an additional buffer and exported after the vivification took place. 
The disadvantage is that vivified clauses are not updated in the exporting 
thread's database.
In this configuration TopoSAT2 does not use a lazy clause exchange 
policy~\cite{audemard_lazy} meaning, clauses are directly exported without any 
additional filter criteria such as times used. 
TopoSAT2 uses a diversification approach as branching, restart, and database 
heuristic, where each solver instance 
uses a different combination of VSIDS or LRB 
branching~\cite{moskewicz_chaff,liang_learning},
dynamic~\cite{audemard_refining}, luby~\cite{een_extensible}, or 
inner-outer restarts~\cite{biere_picosat} and 
Chanseok Oh~\cite{chanseok_between} or Glucose based clause database 
management~\cite{audemard_glucose}. In our experiments, we use 
a newer version\footnote{available on 
\url{https://github.com/the-kiel/TopoSAT2}} than submitted to the SAT 
Competition '18 and reduced the maximum LBD value for vivifying clauses
to 3 for better results on the used system.

\paragraph{Sticky} is an experimental SAT solver based on Glucose using 
physical clause sharing\footnote{available on 
\url{https://github.com/marchartung/sticky}}. Instead of copying clauses during 
export, solvers simply share the reference of a clause, which is used
by the other solvers to access the clause. Therefore, the watcher scheme of 
Glucose was 
adjusted to support such concurrent usage of clauses by ensuring that clauses 
remain constant during propagation. Each clause contains a reference counter to 
keep track of the number of solvers allocating the clause and a placeholder for 
an additional clause reference, in case the clause is replaced.
ECM is used in the same way as in our Glucose implementations except 
that no extra memory for linking the clauses has to be reserved and clauses can 
be exported directly without delay, since improvements are trivially shared due 
to physical clause sharing. The LPCM implementation is different from 
Syrup-LPCM. Sticky does not separate between learned and imported clauses. 
Thus, vivification is also applied to imported clauses. PCM is not implemented 
in Sticky, since physical clause sharing would have to be disabled for 
vivified clauses, which contradicts the main purpose of the solver.

\begin{table}
\centering
\caption{Results for main track benchmarks of SAT Competition '18 (SAT'18), '17 
(SAT'17) and application track of SAT Competition '16 (SAT'16A) with 
15,000 seconds time to solve each of the 1,050 benchmarks.}
\label{tab:complete}\setlength\tabcolsep{0.75ex}
\begin{tabular}{l|rcr|rcr|rcr||rcr} & \multicolumn{3}{c|}{SAT'16A} & 
\multicolumn{3}{c|}{SAT'17} & \multicolumn{3}{c||}{SAT'18} & 
\multicolumn{3}{c}{All}\\
 Solver         & {\smaller SAT} & {\smaller UNSAT} & {\smaller ALL} & 
{\smaller SAT} & {\smaller UNSAT} & {\smaller ALL} & {\smaller SAT} & {\smaller 
UNSAT} & {\smaller ALL} & {\smaller SAT} & {\smaller UNSAT} & {\smaller ALL}\\
 \hline
 \hline
 Syrup          & 77 & 113 & 190 & \textbf{105} & 120 & 225 & 151 & 114 & 265 & 
333 & 
347 & 680\\
 Syrup-PCM      & \textbf{78} & 115 & 193 & \textbf{105} & 120 & 225 & 
\textbf{160} & \textbf{120} & \textbf{\underline{280}} & \textbf{343} & 
355 & \textbf{698}\\
 Syrup-LPCM     & 77 & 116 & 193 & 104 & 120 & 224 & 150 & 119 & 269 & 331 & 
355 & 686\\
 Syrup-ECM3     & 76 & \textbf{122} & \textbf{\underline{198}} & 101 & 
\textbf{126} & 
\textbf{227} & 149 & 119 & 268 & 326 & 
\textbf{367} & 693\\
 Syrup-ECM4     & 72 & 108 & 180 & 98 & 109 & 207 & 133 & 103 & 236 & 303 & 320 
& 623\\
 \hline
 \hline
 Sticky         & 63 &  93 & 156 & \textbf{97} & 109 & 206 & \textbf{143} & 105 
& \textbf{248} & \textbf{303} & 307 & 610 \\
 Sticky-LPCM    & \textbf{68} & \textbf{104} & \textbf{172} & 95 & 117 & 212 & 
135 & 112 & 247 & 298 & \textbf{333} & \textbf{631} \\
 Sticky-ECM3    & 66 & 102 & 168 & \textbf{97} & \textbf{118} & \textbf{215} & 
133 & \textbf{113} & 246 & 296 & \textbf{333} & 629 \\
 \hline
 \hline
 TopoSAT2       & \textbf{80} & \textbf{116} & \textbf{196} & 116 & 
\textbf{122} & \textbf{\underline{238}} & \textbf{161} & \textbf{106} & 
\textbf{267} & 
\textbf{357} & 
\textbf{344} & \textbf{\underline{701}}\\
 TopoSAT2-ECM3  & 75 & 109 & 184 & \textbf{119} & 113 & 232 & 159 & 104 & 263 & 
353 & 
326 & 679\\
\end{tabular}
\end{table}

VSIDS is used as branching heuristic in Sticky. Restart and clause
management heuristics are combinations of luby or dynamic restart and
Chanseok Oh~\cite{chanseok_between} or Glucose based database
management~\cite{audemard_glucose}. During database reduction, solvers
do not distinguish between imported or own clauses, but LPCM is---as
described above---only applied to clauses, which were learned by the
minimizing solver instance. Sticky uses Glucose's lazy clause import
policy~\cite{audemard_lazy} in a relaxed way.  Thereby, every clause
with an LBD smaller than 4 is directly two-watched. For brevity, we
excluded ECM results with a maximal vivification LBD of 4, since they
show a similar performance decrease as Syrup-ECM4.\\

Table~\ref{tab:complete} shows the number of solved instances of the SAT 
Competition'16 (application track), '17 and '18 (main track) by each solver.
TopoSAT2 outperforms the default Syrup implementation in all three 
competitions. Through incorporating vivification, Syrup is able to
solve 2 more 
instances in the SAT Competition'16 (ECM3), 13 more in '18 (PCM) and solves 
over all instances only 3 instances fewer (PCM) than TopoSAT2.

In contrast, the ECM implementation of TopoSAT2 decreases its performance. 
One reason might be the higher vivification overhead compared to Syrup-ECM3, 
since TopoSat2 does not use a lazy export policy, wherefore more clauses are 
vivified and exported. The more likely reason is that vivifications are not 
visible for the learning solver, which decreases the impact of the vivification 
process. Sticky increases the number of solved instances through ECM3 and LPCM 
by around 20 instances, which is the highest increase among all solvers. While 
the number of solved UNSAT instances raised, the number of SAT instances 
decreased, which is similar to the behavior of Syrup-LPCM and Syrup-ECM3. 

\section{Conclusion}
We described multiple homogeneous and parallel learned clause minimization 
approaches based on vivification. Each approach was implemented in the 
state-of-the-art SAT solver Glucose and further tested using SAT problems 
taken from multiple SAT Competitions. Insights gained from these 
implementations are backed by other solver implementations. Our main 
insights are: Applying the basic LCM approach to non-imported clauses (PCM) 
leads to more solved instances in general. Using LCM for shortening a 
smaller part of the exported clauses with a small LBD value (ECM) significantly 
increases the performance on unsatisfiable SAT problems.

\section*{Acknowledgment}
The compute time was provided by the \textit{Norddeutscher Verbund f\"ur Hoch- 
und H\"ochstleistungs\-rechnen} (HLRN). We thank Gilles Audemard and Laurent 
Simon for providing the Glucose implementation as well as Thorsten Ehlers for 
the updated TopoSAT2 implementation.

\bibliographystyle{plain}
\bibliography{REF}

\end{document}